\documentclass[aps,twocolumn,prb]{revtex4-1}
\usepackage{amsmath,amssymb,amsmath}
\usepackage{esint,color}
\usepackage{graphicx}

\newcommand{\epsS}{\varepsilon_{S}}

\newcommand{\Jp}{{\bf J}^d}
\newcommand{\M}[2]{\mathbf{M}_{#1}^{\left(1\right)} \left( r_{#2}  \tilde{\mathbf{r}} \right)}
\newcommand{\N}[2]{\mathbf{N}_{#1} ^{\left(1\right)}\left( r_{#2}  \tilde{\mathbf{r}} \right)}

\newcommand{\rb}{\mathbf{r}}
\newcommand{\rt}{\tilde{\mathbf{r}}}
\newcommand{\rbt}{\tilde{\mathbf{r}}}
\newcommand{\Jr}{\mathbf{J}\left( \mathbf{r'} \right)}
\newcommand{\g}{g \left( \mathbf{r} - \mathbf{r}'\right)}
\newcommand{\go}{g_0 \left( \tilde{\mathbf{r}} - \tilde{\mathbf{r}}'\right)}
\newcommand{\n}{\hat{\mathbf{n}}}
\newcommand{\Ei}{\mathbf{E}_{inc} \left( \mathbf{r} \right)}
\newcommand{\W}{ \mathbf{W} \left( \tilde{\mathbf{r}}' \right)}
\newcommand{\Wn}{ W_n \left( \tilde{\mathbf{r}}' \right)}
\newcommand{\tnabla}{\tilde{\nabla}}
\newcommand{\Who}{ \mathbf{W}_h^\perp }
\newcommand{\Whp}{ \mathbf{W}_h^\parallel }
\newcommand{\Wo}{ \mathbf{W}^\perp }
\newcommand{\ghp}{ \gamma_h^\parallel }
\newcommand{\gho}{ \gamma_h^\perp }
\newcommand{\gao}{ \gamma^\perp }
\newcommand{\Aho}{ \mathbf{A}_h^\perp }
\newcommand{\Ao}{ \mathbf{A}^\perp }
\newcommand{\tOmega}{ \tilde{\Omega} }
\newcommand{\tPi}{ \tilde{\Pi} }
\begin{document}
\title{Magnetoquasistatic Resonances of Small Dielectric Objects}

\author{Carlo Forestiere}
\affiliation{ Department of Electrical Engineering and Information Technology, Universit\`{a} degli Studi di Napoli Federico II, via Claudio 21,
 Napoli, 80125, Italy}
\author{Giovanni Miano}
\affiliation{ Department of Electrical Engineering and Information Technology, Universit\`{a} degli Studi di Napoli Federico II, via Claudio 21,
 Napoli, 80125, Italy}
  \author{Mariano Pascale}
\affiliation{ Department of Electrical Engineering and Information Technology, Universit\`{a} degli Studi di Napoli Federico II, via Claudio 21,
 Napoli, 80125, Italy}
 \affiliation{Photonics Initiative, Advanced Science Research Center, City University of New York, New York, NY, USA}
 \author{Guglielmo Rubinacci}
\affiliation{ Department of Electrical Engineering and Information Technology, Universit\`{a} degli Studi di Napoli Federico II, via Claudio 21,
 Napoli, 80125, Italy}
  \author{Antonello Tamburrino}
\affiliation{Department of Electrical and Information Engineering, Universit\`{a} di Cassino e del Lazio Meridionale, Cassino, Italy}
\affiliation{Department of Electrical and Computer Engineering, Michigan State University, East Lansing, MI 48824 USA}
  \author{Roberto Tricarico}
\affiliation{ Department of Electrical Engineering and Information Technology, Universit\`{a} degli Studi di Napoli Federico II, via Claudio 21,
 Napoli, 80125, Italy}
\affiliation{ICFO Institut de Ciències Fotòniques, The Barcelona Institute of Science and Technology, \\ 08860 Castelldefels, Barcelona, Spain}
 \author{Salvatore Ventre}
\affiliation{Department of Electrical and Information Engineering, Universit\`{a} di Cassino e del Lazio Meridionale, Cassino, Italy}

\begin{abstract}
A small dielectric object with positive permittivity may resonate when the free-space wavelength is large in comparison with the object dimensions if the permittivity is sufficiently high. We show that these resonances are described by the magnetoquasistatic approximation of the Maxwell's equations in which the normal component of the displacement current density field vanishes on the surface of the particle. They are associated to values of permittivities and frequencies for which source-free quasistatic magnetic fields exist, which are connected to the eigenvalues of a  magnetostatic integral operator.  We present the general physical properties of magnetoquasistatic resonances in dielectrics with arbitrary shape. They  arise from the interplay between the polarization energy stored in the dielectric and the energy stored in the magnetic field.  Our findings improve the understanding of resonances in high-permittivity dielectric objects and provide a powerful tool that greatly simplifies the analysis and design of high index resonators.
\end{abstract}

\maketitle

It is well established that small metal  objects with negative permittivity may resonate when the free-space wavelength is large in comparison with their dimensions \cite{Mayergoyz:03,mayergoyz2005electrostatic}.  These resonances can be predicted by the electroquasistatic approximation of the Maxwell-equations, and they are associated to the values of permittivity for which source-free electrostatic fields exist  \cite{Mayergoyz:03,mayergoyz2005electrostatic}.

Small dielectric objects with positive permittivity may also resonate when the free-space wavelength is large in comparison with their dimensions, providing that their permittivity is sufficiently high \cite{VanBladel:75,hulst1981light,Videen}. 
At microwaves, low-loss dielectric materials with relative permittivities $\varepsilon_R$ of the order of $\approx 100$ are routinely used for various applications including resonators and filters \cite{cava2001dielectric,mongia1994low,Reaney06}, while the relative permittivity of certain titanates \cite{schlicke1953quasi,PhysRev.135.A748,sethares1966design} can reach values higher than 1000.

 In the visible and NIR spectral ranges,  resonances in nanoscale high-permittivity objects, such as AlGaAs, Si, Ge nanoparticles, have been observed experimentally, e.g. \cite{evlyukhin2012demonstration,kuznetsov2012magnetic,kuznetsov2016optically}, and have been exploited for several applications, e.g. 
 \cite{Smirnova:16,Schmidt:12,PhysRevB.85.245432,Feng:16,doi:10.1021/acsphotonics.7b00509,baranov2017modifying,staude2013tailoring}.  
In the nano-optics community, these resonances are known as ``Mie resonances'' (e.g. \cite{kuznetsov2016optically}) and  they are described in framework of the full-Maxwell equations (Mie theory \cite{Bohren1998}, quasi-normal modes \cite{Lalanne:18}, characteristic modes \cite{garbacz1965modal}, material-independent-modes \cite{Forestiere:16,Pascale2019}, etc.). 

However, resonances in dielectric objects have a long history which begins at the dawn of the twentieth century with the work on Debye on natural resonant frequencies of free dielectric spheres \cite{debye1909lichtdruck}. In
 1939,  R. D. Richtmyer showed that suitably shaped objects made of a dielectric material can function as electrical resonators at high frequency  \cite{richtmyer1939dielectric}.  
Since the 1960s, they have been used as  high-Q elements for microwave filters and oscillators designs \cite{kajfez1998dielectric}, and following the work of Long et. al \cite{Long:83} also as antennas.  Dielectric resonators has been traditionally analyzed  by using perfect magnetic wall boundary  conditions (PMW) \cite{schlicke1953quasi,sethares1966design,Mongia:94}. However, because electromagnetic fields do exist beyond the geometrical boundary of the cavity  this condition is unable to   accurately predict  resonances  \cite{yee1965natural,Guillon:77,kajfez1998dielectric}. Many ad-hoc corrections to the PMW conditions have been proposed, including the Cohn model \cite{cohn1968microwave}, where  an idealized waveguide with PMW walls is considered, and the 
{Itoh}-{Rudokas} model \cite{Itoh:77}. Instead, Van Bladel  investigated these resonances without ad-hoc assumptions using an asymptotic expansion of the Maxwell's equation in differential form in terms of the inverse of the index of refraction \cite{VanBladel:75}. Glisson et al. \cite{Glisson:83,Kajfez:84} obtained the resonant frequencies of rotationally symmetric dielectric bodies by searching the frequencies at which the determinant of impedance matrix is zero; they assembled the impedance matrix by discretizing a surface integral formulation of the full-Maxwell equation.

In this manuscript, we show by using an integral formulation that the resonances in high-permittivity small dielectric objects can be  predicted by the magnetoquasistatic approximation  \cite{haus1989electromagnetic} of the Maxwell's equations, in which the normal component of the displacement current density field vanishes on the surface of the object. 
These resonances are associated to values of permittivities and frequencies for which source-free quasistatic magnetic fields exist, and they are in one-to-one correspondence to the eigenvalues of the magnetostatic integral operator relating the vector potential inside the object and the displacement current density induced in the object itself.  By studying this operator, we derive the general physical properties of magnetoquasistatic resonances in small high-permittivity objects of arbitrary shape. They arise from the interplay between the polarization energy stored in the dielectric and the energy stored in the magnetic field. Their resonance frequencies constitute an infinite countable set accumulating at $+\infty$, and, for any given shape of the object, they are inversely proportional to its linear dimension and to the square root of its relative permittivity.  The eigenmodes corresponding to different eigenvalues are orthogonal in the usual sense. We also introduce an {\it a-priori} lower bound for the minimum resonance frequency. From a numerical perspective, this approximated integral formulation gives, in its range of validity, a great simplification  with respect  to full-wave approaches and with respect to asymptotic differential formulations.

\section{Resonances in a Magnetoquasistatic Approximation}
Let us consider a homogeneous and isotropic dielectric object with a bounded arbitrary shape $\Omega$ and relative permittivity $\varepsilon_R$.  We define its characteristic size $l_c$ as a chosen linear dimension of the object, and its size parameter  $x=2 \pi l_c / \lambda$, where $\lambda$ is the vacuum wavelength.
We look for source-free solutions of the Maxwell's equations in  high-permittivity dielectric objects, in the limit $x \rightarrow 0 $ (small object). 
Under these conditions, the  electromagnetic field is primarily determined by the displacement current density field $\Jp$ induced inside the object itself \cite{VanBladel:75}. In the appendix \ref{sec:A}, we show that this assumption is well-founded.
 Therefore,  we look for the values of the parameter $\beta = \left( \omega/c_0 \right) \sqrt{\varepsilon_R}$ for which there exists a non-trivial solution of the source-free magnetoquasistatic problem \cite{haus1989electromagnetic}
\begin{subequations}
\begin{align}
     \label{eq:RotA}
     &\nabla \times {\bf A} = \mu_0 {\bf H}, \\
     \label{eq:RotH}
     &\nabla \times {\bf H} =\Jp,
\end{align}
\end{subequations}
with the constitutive relation
\begin{equation}
    \Jp   = \frac{\beta^2}{\mu_0} {\bf A}  \; \Pi_\Omega,
     \label{eq:J} 
\end{equation}     
where  $\Pi_\Omega$ is the characteristic function on the set $\Omega$, i.e. $ \Pi_\Omega \left( \mathbf{r} \right) = 1$ for $\mathbf{r} \in \Omega$, $0$ otherwise,  $\omega$ is the  angular frequency, and $c_0 = 1/\sqrt{\varepsilon_0 \mu_0}$ is the light velocity in vacuum. The magnetoquasistatic vector potential $\mathbf{A}$ satisfies the Coulomb gauge in $\Omega$ and $\mathbb{R}^3 \backslash \Omega$; $\mathbf{A}$ and the quasistatic magnetic field $\mathbf{H}$ are regular at infinity. Eq. \ref{eq:J} disregards the effects of the displacement current density field in  vacuum.  The continuity of the tangential components of $\mathbf{A}$ and $\mathbf{H}$ imply, respectively, the continuity of the normal components of $\mathbf{H}$ and $\Jp $ across the boundary $\partial \Omega$ of $\Omega$. Since the normal component of the current density field  $\Jp $ at the boundary $\partial \Omega$ is equal to zero, the current density field $\Jp $ is div-free everywhere in $\mathbb{R}^3$; instead, the normal component of the vector potential at  $\partial \Omega$ may be discontinuous. The fact that $\Jp$ has a vanishing normal component on $\partial \Omega$ implies that also the normal component of the polarization current density field is zero.

It is convenient to scale the spatial coordinates by the characteristic size of the object $l_c$, $\mathbf{r} = l_c \tilde{\mathbf{r}}$. Thus, we denote with $\tilde{\Omega}$ the scaled domain $\Omega$.
Then, problem \ref{eq:RotA},\ref{eq:RotH},\ref{eq:J} is solved by expressing the vector potential $\mathbf{A}$ in terms of the current density $\Jp $ as:

\begin{equation}
  \mathbf{A} \left( \tilde{\mathbf{r}} \right) = \mu_0 l_c^2 \mathcal{L}_m \left\{ \Jp  \right\} \left( \tilde{\bf r} \right),
 \label{eq:AJ}
\end{equation}
where we have introduced the magnetostatic integral operator
\begin{equation}
\mathcal{L}_m \left\{ \Jp  \right\} \left( \rbt \right) =  \iiint_{\tilde{\Omega}} \Jp  \left( \rbt' \right)  g_0\left(  \rbt - \rbt' \right)  \text{d}\tilde{V} \qquad \forall \rbt \in \tilde{\Omega},
\label{eq:IntegralOperator}
\end{equation}
and  $g_0 \left( \mathbf{r} \right)  = 1/(4\pi r)$ is the static Green function in vacuum. In \ref{eq:IntegralOperator} there is the static Green function because we are neglecting the displacement current density in vacuum. By combining Eqs. \ref{eq:J} and \ref{eq:AJ}, we obtain the linear eigenvalue problem
\begin{equation}
 \Jp \left(  \rbt \right) = y^2 \mathcal{L}_m \left\{ \Jp  \right\} \left( \rbt \right)  \qquad \forall \rbt \in \tilde{\Omega},
\label{eq:Auxiliary}
\end{equation}
where $y = l_c \left( \omega / c_0 \right) \sqrt{\varepsilon_R} = x \sqrt{\varepsilon_R}$. Equation \ref{eq:Auxiliary} holds in the weak form in the functional space constituted by the functions which are div-free within $\tilde{\Omega}$ and having zero normal component on $\partial \tilde{\Omega}$, and equipped with the inner product
$ \langle \mathbf{w},\mathbf{v} \rangle_{\tilde{V}} = \iiint_{\tilde{V}} \mathbf{w}^* \cdot \mathbf{v} \, \mbox{d} \tilde{V}.$

The operator $\mathcal{L}_m$ is compact, positive-definite, and self-adjoint. As a consequence: Equation \ref{eq:Auxiliary}  admits a countable set of eigenvalues $\left\{y_n^2 \right\}_{n \in \mathbb{N}}$ and eigenmodes $\Jp _n$; the eigenvalues $y_{n}^2$ are real and positive and accumulate at infinity. The eigenmodes corresponding to different eigenvalues are orthogonal in the usual sense; and they constitute a complete basis of the considered functional space.
Furthermore, the eigenvalue $y_n$ is proportional to the magnetic energy of the corresponding eigenmode:
\begin{equation}
  y_n = \frac{\left\| \nabla_{\tilde{\mathbf{r}}} \times \mathbf{A}_n \right\|_{\mathbb{R}^3}}{\left\| \mathbf{A}_n \right\|_{\tilde{\Omega}}}
  \label{eq:PropertiesEig2}
\end{equation}
$\mathbf{A}_n$ is the magnetic vector potential generated by $\Jp _n$ in the whole space, and  $\left\| {\bf v} \right\|_{V}^2 = \langle {\bf v}, {\bf v} \rangle_{V}$.  

The resonance angular  frequencies $\omega_n$ are given by
\begin{equation}
	\omega_n = \frac{c_0}{l_c \sqrt{\varepsilon_R}} y_n.
   \label{eq:ResonantFreq}
\end{equation}
The mathematical structure of the integral operator \ref{eq:IntegralOperator} does not depend on the linear characteristic dimension of the dielectric object $l_c$, namely it is scale invariant. This fact combined with Eq. \ref{eq:ResonantFreq} leads to the important property of the magnetoquasistatic resonances: for any given shape of the object the resonance frequencies are always inversely proportional to both $l_c$ and $\sqrt{\varepsilon_R}$. Furthermore, the resonance frequencies accumulate at infinity. By contrast, we recall that in small metal particles the electrostatic resonance frequencies accumulate at finite frequencies \cite{Mayergoyz05} (for arbitrarily shaped particles with a simple Drude dispersion they accumulate at $\omega_p/\sqrt{2}$ where  $\omega_p$ is the plasma frequency of the metal.)

Moreover, the following bound on the eigenvalues hold:
\begin{equation}
  \label{eq:Bound1}
  y_n  \ge \frac{\sqrt[4]{3}}{2  \sqrt{ \pi }} \frac{1}{ \tilde{a}} \approx \frac{0.37}{\tilde{a}} \qquad \forall n \in \mathbb{N} 
\end{equation}
where $\tilde{a}$ is the radius of a sphere $\tilde{B}_a$ having the same volume of $\tilde{\Omega}$.
The inequality \ref{eq:Bound1} is obtained by multiplying both members of Eq.  \ref{eq:Auxiliary} by $\Jp$, integrating over $\tilde{\Omega}$ and using the Cauchy-Schwarz inequality and the inequality \cite{jeffreys1999methods}
\begin{equation}
\begin{aligned}
\int_{\tilde{\Omega}} \frac{1}{\left| {\bf r} - {\bf r'} \right|^2} d \tilde{V}'
&\le  \int_{\tilde{B}_a} \frac{1}{\left| {\bf r} - {\bf r'} \right|^2} d \tilde{V}' = 4 \pi \tilde{a} && \forall {\bf r} \in \tilde{\Omega}  \\
 \int_{\tilde{\Omega}}  \left| \Jp \right| d \tilde{V}' & \le \sqrt{ \text{mis} \left\{ \tilde{\Omega} \right\}} \sqrt{ \int_{\tilde{\Omega}}  \left| \Jp \right|^2 d \tilde{V}'}
 \end{aligned}
\end{equation}
where  $\tilde{B}_a$ is centered in ${\bf r}$.
Let us apply the bound \ref{eq:Bound1} to the case of an oblate spheroid: by keeping the major semi-axis fixed, and decreasing the minor one, the volume of the particle decreases and so does $\tilde{a}$: this leads to an increase of lower bound for the eigenvalues (and for the magnetostatic energies) of all the modes. This is in contrast with what it is observed for a metal spheroid, where the same decrease of the minor semi-axis implies a {\it decrease} of the electrostatic energy of the fundamental electrostatic (plasmon) mode \cite{maier2007plasmonics}.

In a ``material resonance picture'' \cite{Forestiere:16,Forestiere:19}, once the operating frequency is assigned, the resonance permittivities are given by:
\begin{equation}
	\varepsilon_{R,n} = \left( \frac{y_n}{x} \right)^2.
\end{equation}
Since $\mathcal{L}_m$ is positive-definite, source-free  magnetoquasistatic field may exist only for positive permittivities. Moreover, for any given shape, they $\varepsilon_{R,n}$ are inversely proportional to $x^2$, they also constitute accumulate at $\infty$. It is now apparent the difference between the magnetoquasistatic resonances and the plasmon resonances, which only exist for negative permittivities, are size-independent, and  accumulate at $-1$ \cite{mayergoyz2005electrostatic}.

In the Appendix \ref{sec:A}, we  show that in the limit $x \rightarrow 0$ the Maxwell equations admit two orthogonal sets of current modes. The  modes of the first set are  div-free and curl-free within the object,  but have a non-vanishing normal component on the object surface.   The  modes of the second set  are div-free  within the object,  have a vanishing normal component on the object surface and non-zero curl. The latter set is solution of the eigenvalue problem \ref{eq:Auxiliary}  and the corresponding resonance frequencies $\omega_n$ are given by \ref{eq:ResonantFreq}.

In a magnetoquasistatic resonance the energy oscillates back and forth between the polarization energy stored in the dielectric and the energy stored in the magnetic field, as shown in Appendix \ref{sec:B}.

The eigenvalue problem \ref{eq:Auxiliary} can be numerically solved through a finite element  approach, briefly outlined in Appendix \ref{sec:C}, where, unlike differential formulations, only the spatial domain $\Omega$ is discretized and the radiation condition are automatically satisfied. This approach only requires the calculation of the eigenvalues of a real symmetric matrices, for which efficient routines exists \cite{van1983matrix,laug}.

The magnetostatic formulation can be easily extended to the scenario where the object is standing on a substrate with relative permittivity $\epsS$, which is very relevant for the applications. Since the thickness of the substrate is typically much larger than the dimensions of the object, we can safely assume it to be semi-infinite. Therefore, by using the method of images \cite{jackson1999classical}, the resonance are still associated to the eigenvalues of the operator \ref{eq:IntegralOperator} provided that the Green function $g_0$ in Eq. \ref{eq:IntegralOperator} is replaced with
\begin{equation}
	g_{S} \left( \rb, \rb' \right)  = \frac{1}{\left| \rb - \rb' \right|} - \frac{\epsS-1}{\epsS+1}\frac{1}{\left|\rb - \rb''\right|}
\end{equation}
where $\mathbf{r}''$ is the mirror image of $\mathbf{r}'$ with respect to the substrate plane.

\section{Results and Discussions}
\begin{figure}[!ht]
\centering
\includegraphics[width=0.95\columnwidth]{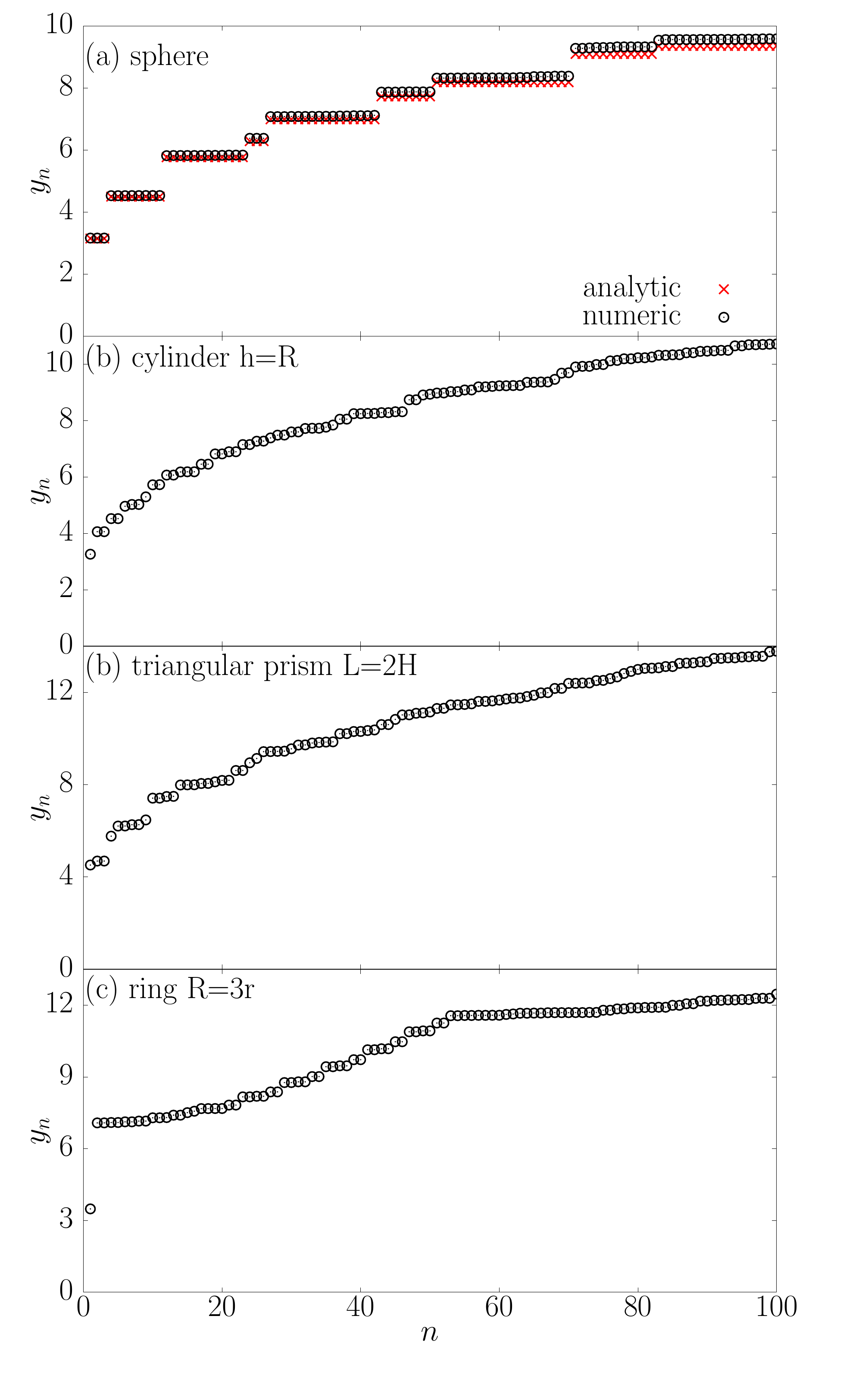}
	\caption{(a) Comparison between the  eigenvalues $y_n$ of a sphere and the poles of the Mie coefficients in the limit $x \rightarrow 0$. (b) Eigenvalues of a finite-size cylinder with height $H$ equal to the radius $R$, with $l_c = R$. (c) Eigenvalues of a right triangular prism with height $H$ and edge $R=2H$ with $l_c = R$. (d) Eigenvalues of a ring with minor radius $r$ and major radius  $R=3r$, with $l_c = 3R$}
\label{fig:Eigenvalues}
\end{figure}

{\bf Sphere}. To benchmark the magnetoquasistatic approximation, we now consider a dielectric sphere for which there exists a full-wave analytic solution, i.e. the Mie theory \cite{Bohren1998}.   Van De Hulst gives  \cite{hulst1981light}  the resonant conditions of a high-permittivity small  sphere by finding the poles of the Mie coefficients  in the limit $x \rightarrow 0$. The resonances occur at $y_n = r_{n,l}$ for the TM multipoles and at $y_n = r_{n-1,l}$ for the TE multipoles, for any $n,l \in \mathbb{N}$, where $r_{n,l}$ denotes the $l$-th zero of the spherical Bessel function $j_n$.   In Fig. \ref{fig:Eigenvalues} (a) we compare the first 100 magnetoquasistatic eigenvalues  of the operator $\mathcal{L}_m$ with the poles of the Mie coefficients, while in Table \ref{tab:ResonancesSiSingle} we show the corresponding  values and the numerical error. The details on the hexahedral mesh used for the computation are given in Appendix \ref{sec:C}. In particular, we note that for the first 50 eigenvalues the numerical error is below $2 \%$. The magnetoquasistatic model correctly predicts the peculiar degeneracies of the TE and TM resonances: the $l^{th}$ resonance of the $n^{th}$ TE multipole is degenerate with the $l^{th}$ resonance of the $\left(n-1 \right)^{th}$ TM multipole: namely the eigenvalue $r_{0,l}$ has a degeneracy 3, while  $r_{n,l}$ with $n\ge 1$ has a degeneracy $4(n+1)$.

\begin{table}[h!]
\centering
\begin{tabular}{|c|c|c|c|c|c|c|c|c|c|}
\hline
$\#$ & 1-3 & 4-11 & 12-23 & 24-26 & 27-32 & 33-36  & 37-38 & 39-42 & 43-50 \\
\hline
$y_n$ & 3.16  & 4.53 &  5.82 & 6.38 & 7.07 & 7.08  & 7.09 & 7.10 & 7.87  \\
\hline
$r_{nl}$ & 3.14 & 4.49 & 5.76 & 6.28 & 6.99 & 6.99 & 6.99 & 6.99 & 7.73 \\
\hline
$\epsilon \left[ \% \right]$ & 0.64  & 0.89 & 1.04 & 1.59 & 1.14 & 1.29 & 1.43 & 1.57 & 1.81 \\
\hline

\end{tabular}
\caption{Eigenvalues of a sphere $y_n$ compared with the poles of the Mie coefficients $r_{nl}$ in the limit $x \rightarrow 0$. Relative error $\epsilon$.}
\label{tab:ResonancesSiSingle}
\end{table}

\begin{figure*}[!ht]
\centering
\includegraphics[width=0.7 \textwidth]{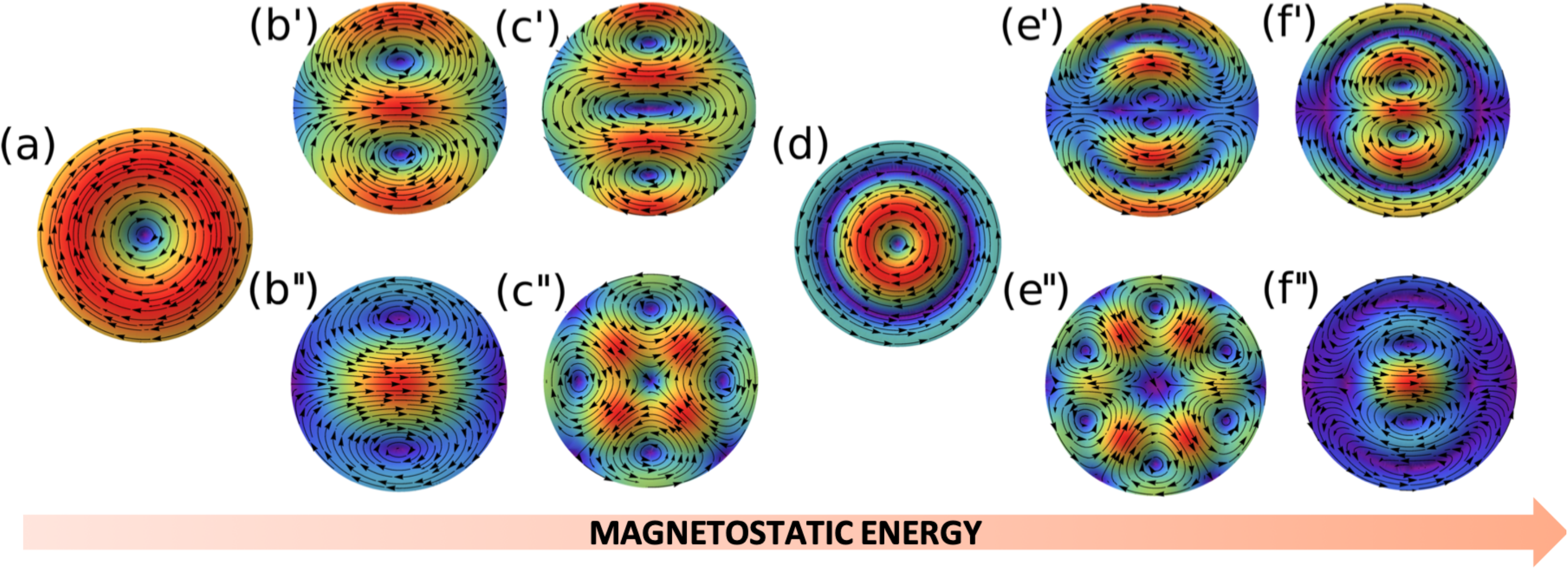}
\caption{Eigenmodes of a sphere, associated to the first 6 distinct eigenvalues $y_n$. The modes on the same column are degenerate. We represent the projection of the modes on the plane $y=0$ in panels (a),(b''),(c''),(d),(e''),(f''), on $z=0.4 R$ in (b'), on $z=0.35 R$ in (c'), on $z=0.26 R$ in (e'), on $z=0.23 R$ in (f').}
\label{fig:Modes}
\end{figure*}

The magnetoquasistatic modes are ordered according to  their magnetic energy, which does not necessarily follow their multipolar order.
The first eigenvalue $y = r_{01}$ is associated to three degenerate TE displacement current density modes, $\M{pm1}{01}$ with $p=e,o$ and $m=0,1$, which correspond to magnetic dipoles oriented along the three coordinate axis. As an example, we show $\M{o11}{01}$  in Fig. \ref{fig:Modes} (a). 
The fact that the lowest-energy  magnetoquasistatic mode of a high-permittivity small sphere is the magnetic dipole is in agreement with experimental studies \cite{kuznetsov2012magnetic}.
The second eigenvalue has an eight fold degeneracy: it is associated to five TE modes $\M{pm2}{11}$ with $p=e,o$ and $m=0-2$ (in Fig. \ref{fig:Modes} (b') $\M{o12}{11}$ is shown) and three TM modes $\N{pm1}{11}$ with $m=0-1$ and $p=e,o$ (in Fig. \ref{fig:Modes} (b'') $\N{e11}{11}$ is shown). The third eigenvalue is associated to seven TE modes $\M{pm3}{21}$ with $p=e,o$ and $m=0-3$ (in Fig. \ref{fig:Modes} (c') $\M{o13}{21}$ is shown) and five TM modes $\N{pm2}{21}$ with $p=e,0$ and $m=0-2$ (in Fig. \ref{fig:Modes} (c'') $\N{e12}{21}$ is shown). The fourth eigenvalue is associated to three TE modes $\M{pm1}{02}$ with $m=0-1$ and $p=e,o$, associated to two counter-rotating current loops (we show $\M{o11}{02}$ in Fig. \ref{fig:Modes} (d)).
The fifth eigenvalue is associated to nine TE modes $\mathbf{M}_{pm4} \left( r_{31} \mathbf{r}' \right)$ with $m=0-4$ and $p=e,o$  (in Fig. \ref{fig:Modes} (e') $\M{o14}{31}$ is shown) and seven TM modes $\N{pm3}{31}$ with $m=0-3$ and $p=e,0$ (in Fig. \ref{fig:Modes} (e'') $\N{e13}{31}$ is shown). The sixth eigenvalue is associated to five TE modes $\M{pm2}{12}$ with $m=0-2$ and $p=e,o$  (in Fig. \ref{fig:Modes} (f') $\M{o12}{12}$ is shown) and three TM modes $\N{pm1}{22}$ with $m=0-3$ and $p=e,0$ (in Fig. \ref{fig:Modes} (f'') $\N{e13}{31}$ is shown). All the magnetoquasistatic modes have vanishing  normal component and non-zero tangential component on the sphere boundary.

\begin{figure}[!ht]
\centering
\includegraphics[width=0.9\columnwidth]{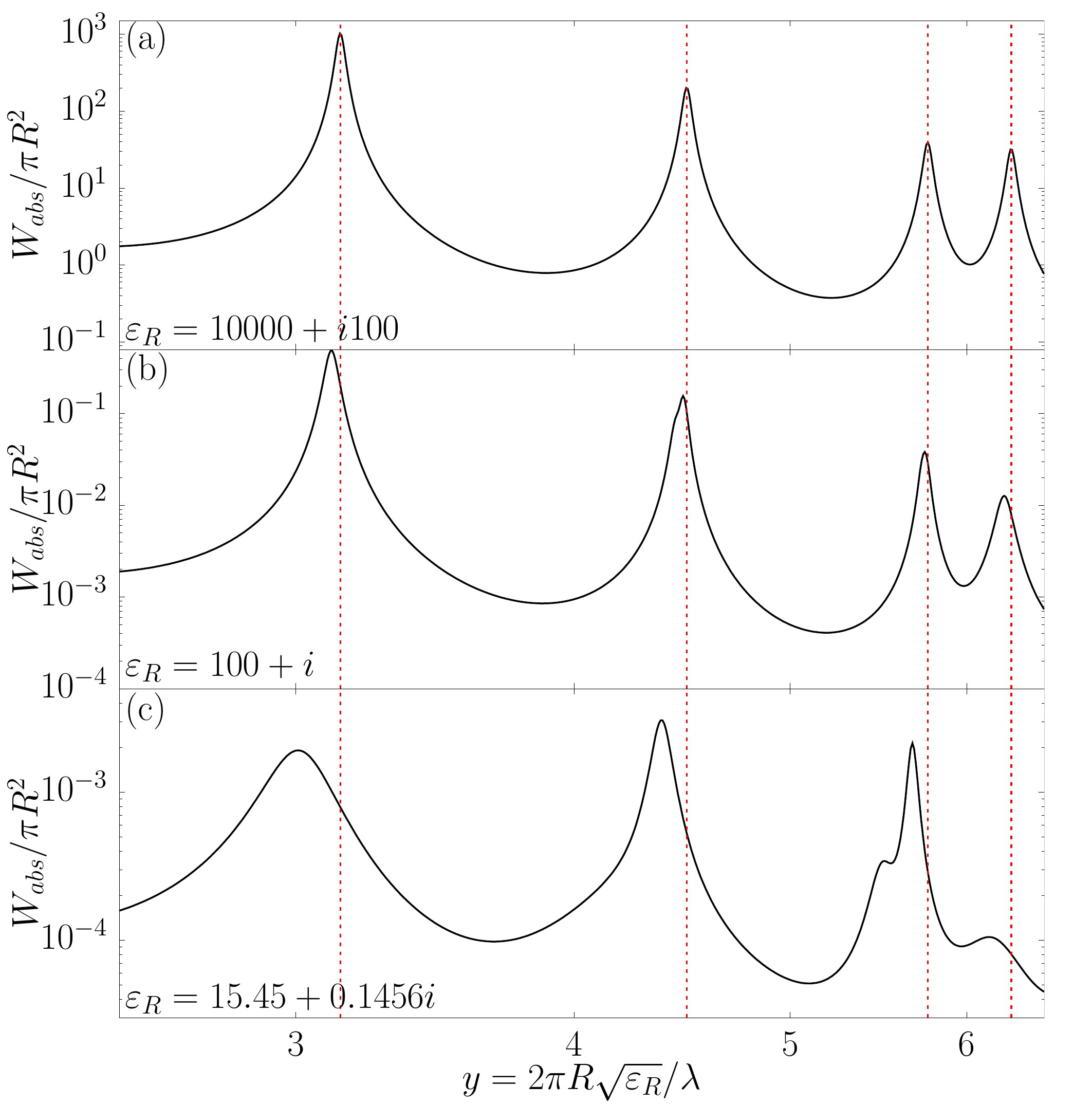}
\caption{Normalized power absorbed by a sphere with radius $R = l_c$ and $\varepsilon_R =$ (a)  $10^4 + 10^2 i $, (b) $10^2 + 1 i $, and (c) $15.45 + 0.1456 i $ as a function of the parameter $y = 2 \pi R \sqrt{\varepsilon_R} /\lambda $. The sphere is centred in $\left( 0, 0, 0 \right)$ and it is excited by an electric dipole $\mathbf{N}_{e11}^{\left(3\right)}$ at position $\left( 0, 0, 1.5 R \right)$. The first four eigenvalues $y_n$ of Tab. \ref{tab:ResonancesSiSingle} are shown with vertical dashed lines.}
\label{fig:PabsSphere}
\end{figure}

We now show that the magnetoquasistatic eigenvalues predict the occurrence of the resonance peaks in the scattering response of a high-permittivity small sphere.  In Fig. \ref{fig:PabsSphere} we show with a continuous line the power absorbed by the sphere normalized by $\pi R^2$  as a function of $y= x \sqrt{\varepsilon_R}$ for different values of $\varepsilon_R$, calculated by the Mie theory \cite{Bohren1998}, and with four vertical dashed lines the first non-degenerate eigenvalues $y_n$ of the magnetostatic volume integral operator listed in Tab \ref{tab:ResonancesSiSingle}. Thus, the first vertical line represents the position of the mode $\M{o11}{01}$ , the second line is associated to one TE mode $\M{o12}{11}$ and one TM mode $\N{e11}{11}$, the third is associate to $\M{o13}{21}$ and $\N{e12}{21}$, the forth line to $\M{o11}{02}$. They are shown in Fig. \ref{fig:Modes} (a-d). In Fig. \ref{fig:PabsSphere} (a) we consider   $x \in \left[0.025,0.065  \right] \ll 1$ and $\varepsilon_R = 10^4 + i 10^2$. We find very good agreement between the eigenvalues $y_n$ and the absorption peaks positions. This is expected because the investigated values of $x$ are much less than one.
Then in Fig. \ref{fig:PabsSphere} (b) we consider  $x \in \left[0.25,0.65  \right] < 1$ and  $\varepsilon_R = 10^2 + i$. We note a red-shift of the peaks with respect to the magnetoquasistatic prediction,  because the values of $x$ starts to be comparable to one. Eventually, in Fig. \ref{fig:PabsSphere} (c) we investigate a Silicon sphere with $x \in \left[0.64, 1.6 \right]$ and $\varepsilon_R = 15.45+0.1456i$. Although $x$ is now comparable to one, there is still a correlation between the peaks and the magnetoquasistatic predictions. However, the peaks now show a broadening and a red-shift.  The third and fourth peaks appearing in Fig.  \ref{fig:PabsSphere} (c)  arise from the splitting of the third peak of panel (b) because  the degeneracy of the modes  $\M{o13}{21}$ and $\N{e12}{21}$ is lifted by the retardation.

{\bf Finite-Size Cylinder.} We now investigate a finite-size cylinder of radius $R$ and height $h=R$, which is very common among nanofabricated structures, because it is compatible with planar nanofabrication processes.   Although no analytic solution exists in this case,  semi-empirical formulas have been proposed for low-index resonances obtained by brute-force numerical calculations \cite{Mongia:94}. We assume a characteristic size $l_c=R$. 
\begin{figure*}[!t]
\centering
\includegraphics[width=1\textwidth]{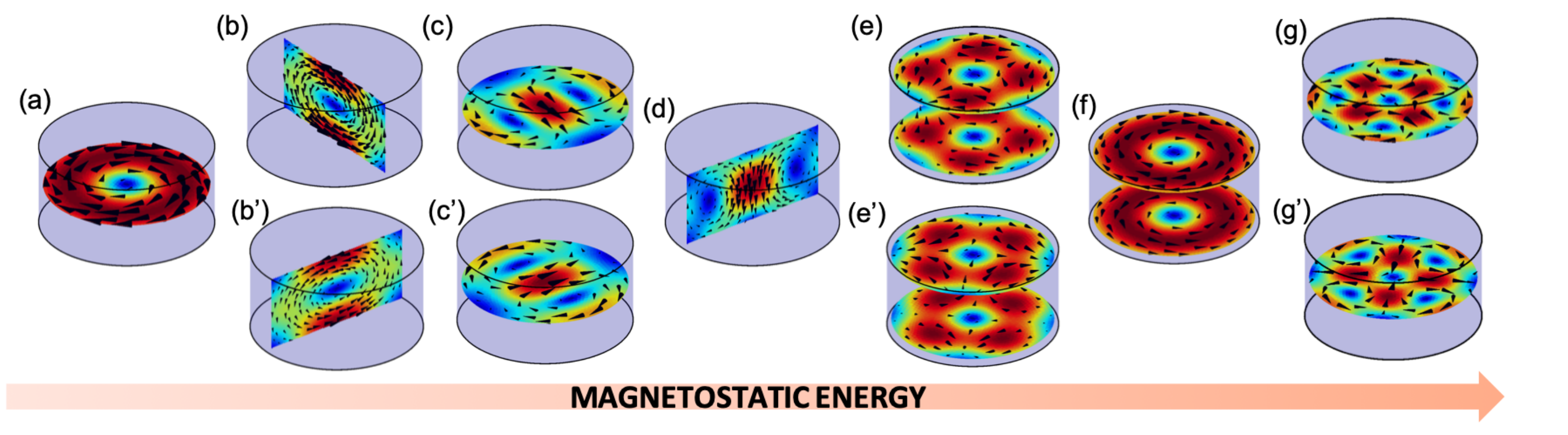}
\caption{Eigenmodes of a  finite-size cylinder with height equal to the radius, associated to the first 4 distinct eigenvalues $y_n$. The modes on the same column are degenerate. (a) $\text{TE}_{01\delta}$, (b) $\text{HEM}_{11\delta}$, (c) $\text{HEM}_{12\delta}$, (d)  $\text{TM}_{01\delta}$, (e)  $\text{HEM}_{21\delta}$, (f) $\text{HEM}_{011+\delta}$, (g) $\text{HEM}_{22\delta}$,}
\label{fig:ModesCylinder}
\end{figure*}
In Fig. \ref{fig:Eigenvalues} (b) we show the first $100$ magnetoquasistatic eigenvalues of the investigated cylinder.  
In Fig. \ref{fig:ModesCylinder} we plot the displacement current density modes corresponding to the first seven distinct  eigenvalues.  A similar {\it catalog} has been produced by Kajfez et al. \cite{Kajfez:84}  by using a surface integral equation formulation of the full-Maxwell equations for bodies of revolution. Here, we follow the classification introduced of Glisson et al. \cite{Glisson:83}, which is in turn borrowed from the literature on cylindrical waveguides \cite{snitzer1961cylindrical,Standards}. Specifically, the modes which are symmetric along the azimuthal direction are denoted either as $\text{TM}_{0n\delta}$ or as $\text{TM}_{0n\delta}$. All the remaining modes are known as {\it hybrid} modes with respect to the axis of rotation, and they are denoted by $\text{HEM}_{mnp}$, where the subscripts $m$, $n$, $p$ are associated to the number of oscillation of the mode along the azimuthal, radial, and axial directions. It is also worth to note the third subscript is denoted as $\delta$ if smaller than unity. 
The fundamental mode of the finite size cylinder shown in Fig. \ref{fig:ModesCylinder} (a) is $\text{TE}_{01\delta}$, i.e. a magnetic dipole oriented along the vertical axis. The next two degenerate modes, e.g. $\text{HEM}'_{11\delta}$- $\text{HEM}''_{11\delta}$, shown in  Fig. \ref{fig:ModesCylinder} (b')-(b''), radiate like magnetic dipoles oriented along two orthogonal horizontal directions.  We note that while in a high-permittivity small sphere the three magnetic dipoles are degenerate, in the investigated cylinder they are split up by the symmetry breaking, with $\text{TE}_{01\delta}$ having lower magnetoquasistatic energy with respect to $\text{HEM}'_{11\delta}$- $\text{HEM}''_{11\delta}$.  Moving further in the direction of increasing magnetostatic energy, we encounter the hybrid modes $\text{HEM}'_{12\delta}$,$\text{HEM}''_{12\delta}$ followed by the azimuthally symmetric mode $\text{TM}_{01\delta}$. Then, we encounter the hybrid modes $\text{HEM}_{21\delta}'$, $\text{HEM}_{21\delta}''$, then $\text{TE}_{011+\delta}$, followed by the modes $\text{HEM}_{22\delta}'$,$\text{HEM}_{22\delta}''$.    
 
 In Table \ref{tab:MQScyl} we compare the values of the magnetoquasistatic eigenvalues $y_n$ against semi-empirical formulas \cite{Mongia:94} (when available) obtained by fitting the results of different numerical methods for the modes $\text{TE}_{01\delta}$ \cite{DeSmedt:84},  $\text{HEM}_{11\delta}$ \cite{kishk1993computed}, $\text{TM}_{01\delta}$ \cite{tsuji1984analytical},  $\text{TE}_{011+\delta}$  \cite{DeSmedt:84}.

\begin{table}[ht!]
\centering
\begin{tabular}{|c||c|c|c|c|c|c|}
    \hline
	$\#$ & 1 & 2-3 & 4-5 & 6 & 7-8 & 9  \\
    \hline 
	name  & $\text{TE}_{01\delta}$  &  $\text{HEM}_{11\delta}$ & $\text{HEM}_{12\delta}$ & $\text{TM}_{01\delta}$ & $\text{HEM}_{21\delta}$ & $\text{TE}_{011+\delta}$  \\
	\hline
	$y_n$ & 3.26 & 4.05 &  4.52 & 4.96 & 5.02 & 5.30 \\
	\hline
	$ \tilde{y}_n $ & 3.22 & 4.11 &  - & 4.95 & - & 5.28  \\
\hline
\end{tabular}
\caption{Magnetostatic eigenvalues $y_n$ for an isolated finite-size cylinder with height equal to the radius. $\tilde{y}_n$ are the corresponding values evaluated by the empirical approximated formulas  \cite{Mongia:94} (if available).}
\label{tab:MQScyl}
\end{table}

\begin{figure}[!ht]
\centering
\includegraphics[width=0.9\columnwidth]{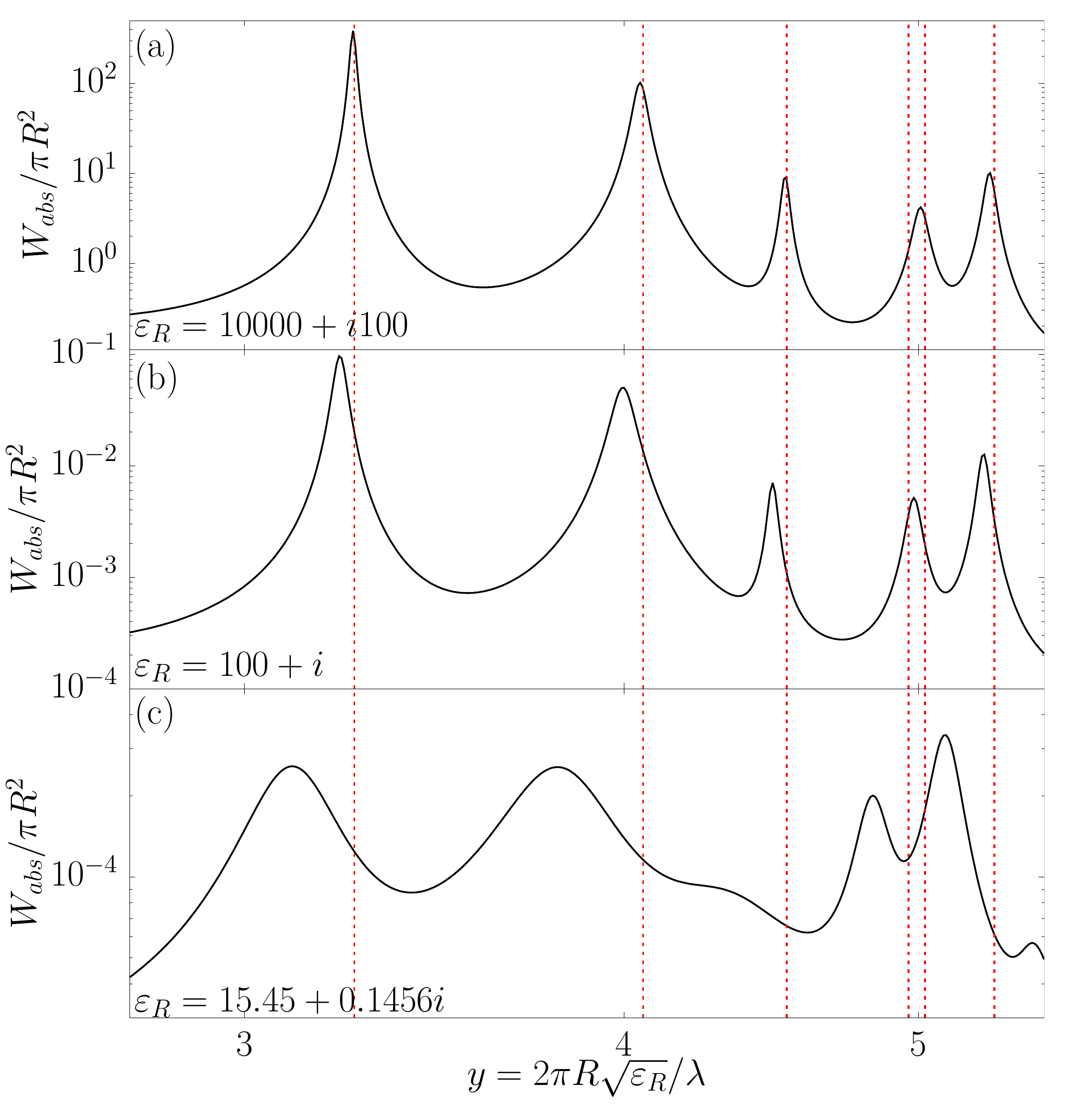}
\caption{Normalized power absorbed by a cylinder with radius $R = l_c$, height $h=R$, and $\varepsilon_R=$  (a) $10^4 + 10^2 i $, (b) $10^2 + 1 i $, (c) $15.45 + 0.1456 i $, as a function of $y = x \sqrt{\varepsilon_R}$. The cylinder is centered in $\left( 0, 0, 0 \right)$ and it is excited by an electric dipole $\mathbf{N}_{o11}^{\left(3\right)}$ at position $\left( R, 0, 1.5 R \right)$. The first six magnetoquasistatic eigenvalues $y_n$ are shown with vertical dashed lines.}
\label{fig:PabsCylinder}
\end{figure}

As for the sphere, we show in Fig. \ref{fig:PabsCylinder} that from the knowledge of the magnetoquasistatic eigenvalues we can predict the occurrence of the resonance peaks in the absorbed power spectra. The absorbed power has been calculated by an independent full-wave numerical method, i.e. the PMWCHT approach \cite{harrington1993field,ForestiereSIE}.  We show with vertical dashed lines the first six non-degenerate eigenvalues $y_n$   associated to the modes $\text{TE}_{01\delta}$, $\text{HEM}_{11\delta}$, $\text{HEM}_{12\delta}$, $\text{TM}_{01\delta}$, $\text{HEM}_{21\delta}$, $\text{TE}_{011+\delta}$, whose values are listed in Tab. \ref{tab:MQScyl}. In Fig. \ref{fig:PabsCylinder} (a) we consider $\varepsilon_R=10^4 +10^2i$ and $x \in \left[0.025,0.065  \right] \ll 1$:  
The magnetoquasistatic eigenvalues exactly predict the absorption peaks because the hypotheses of the magnetoquasistatic model are verified.  Next, in Fig. \ref{fig:PabsCylinder} (b) we consider $x \in \left[0.25,0.65  \right] < 1$ and $\varepsilon_R=10^2 +10i$:  we note a red-shift and a broadening of the peaks  because the values of $x$ are approaching one. Eventually, in Fig \ref{fig:PabsCylinder} (c), we consider a silicon cylinder with $x \in \left[0.64, 1.6 \right] \approx 1$ and $\varepsilon_R = 15.45 + 0.1456 i$. Even if $x \approx 1$, there is still a correlation between the peaks and the magnetoquasistatic predictions. However, the shift and the broadening of the peaks are now significant.

\begin{figure*}[!t]
\centering
\includegraphics[width=0.84\textwidth]{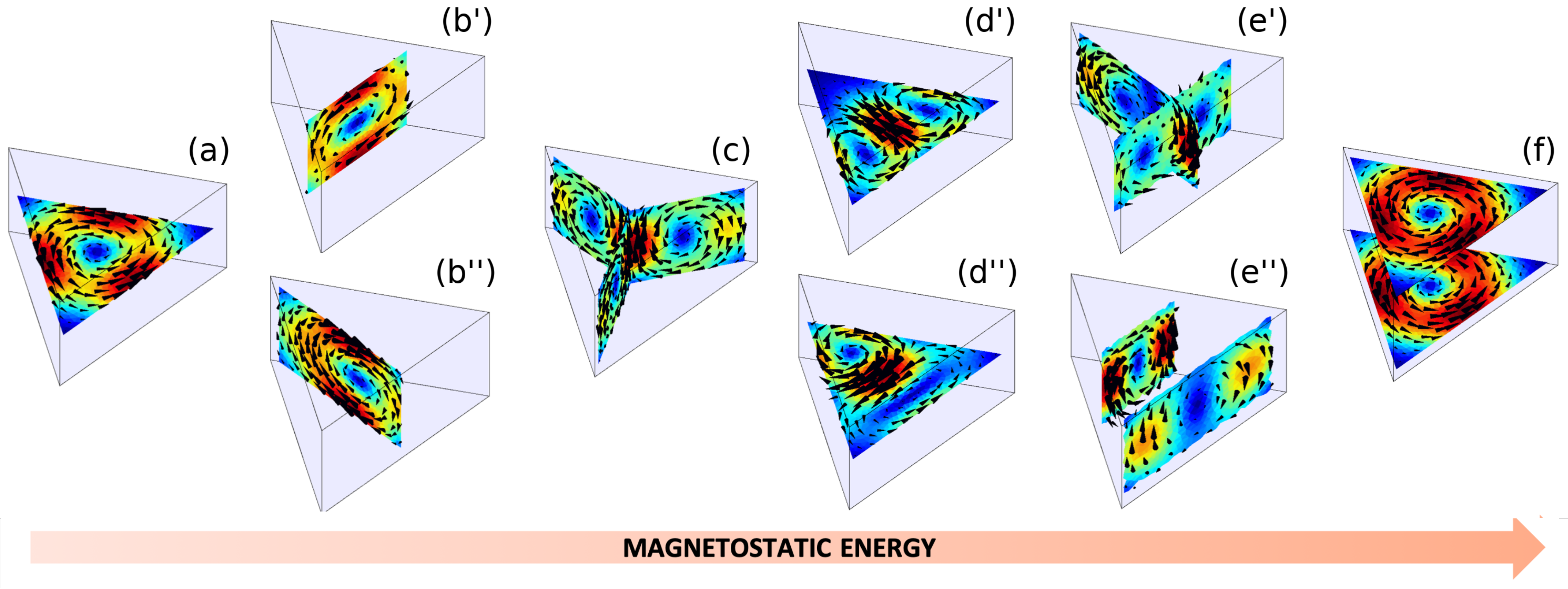}
\caption{Eigenmodes of a triangular prism with $L=2H$ associated to the first six non-degenerate eigenvalues.}
\label{fig:ModesTri}
\end{figure*}

\begin{table}[ht!]
\centering
\begin{tabular}{|c||c|c|c|c|c|c|c|}
    \hline
	$\#$ & 1 & 2-3 & 4 & 5-6 & 7-8 & 9 & 10-11  \\
    \hline 
	$y_n$ & 4.52 & 4.69 &  5.76 & 6.21 & 6.26 & 6.48 & 7.41  \\
\hline
\end{tabular}
\caption{Magnetoquasistatic eigenvalues $y_n$ of an isolated triangular prism with edge $L$ and height $H$, with $L=2H$. We assumed $l_c = L$.}
\label{tab:MQStri}
\end{table}

{\bf Right Triangular Prism.}  We now present the magnetoquasistatic analysis of a right triangular prism, which is not axisymmetric as the objects investigated so far, but belongs to the symmetry group $D_{3h}$.
The triangular prism has height $H$, while its basis is an equilater triangle of edge $L=2H$. In Fig. \ref{fig:Eigenvalues} (c) we show the first 100 magnetostatic eigenvalues, while in Fig. \ref{fig:ModesTri} we show the corresponding modes. It is apparent from Tab. \ref{tab:MQStri} that the eigenvalues have multiplicity one or two, which is consistent with the fact that the prism is invariant under transformation of the group $D_{3h}$  \cite{lyubarskii2013application}. 
The fundamental mode of a triangular prism is a magnetic dipole oriented along the vertical axis, shown in Fig.  \ref{fig:ModesTri} (a). The next two degenerate modes are magnetic dipoles lying on the horizontal plane and oriented  along one height and the corresponding orthogonal edge, as shown in Fig.  \ref{fig:ModesTri} (a)-(b). 
By moving further in the direction of increasing magnetostatic energy we note that the modes in Fig. \ref{fig:ModesTri} (c) and (d)-(d') resemble the modes $\text{TM}_{01\delta}$ and $\text{HEM}_{12\delta}$ of the cylinder but the order of the corresponding magnetostatic energy is inverted. This fact suggests that the order of the magnetoquasistatic resonances can be tailored by a convenient design of the geometry of the object.

\begin{figure*}[!ht]
\centering
\includegraphics[width= \textwidth]{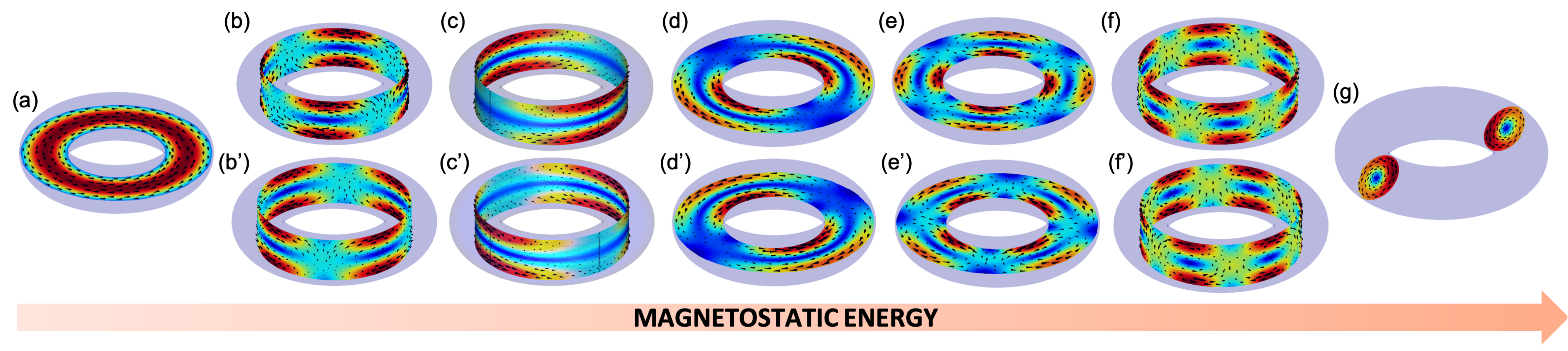}
\caption{Eigenmodes of a ring, associated to the first 7 distinct eigenvalues $y_n$. The modes on the same column are degenerate. We represent the projection of the modes on the intersection of the ring with the plane $z=0$ in panels (a),(d),(d'),(e),(e'); with the surface $x^2+y^2=R^2$ in (b),(b'), (c),(c'),  (f),(f'); with the plane $x=0$ in (g).}
\label{fig:ModesTorus}
\end{figure*}

\begin{table}[ht!]
\centering
\begin{tabular}{|c||c|c|c|c|c|c|c|}
    \hline
	$\#$ & 1 & 2-3 & 4-5 & 6-7 & 8-9 & 10-11 & 12   \\
    \hline
	$\#$ & $\text{T}_{010}$ & $\text{HTP}_{211}^\updownarrow$ & $\text{HTP}_{111}^\updownarrow$ & $\text{HTP}_{111}^\leftrightarrow$ & $\text{HTP}_{211}^\leftrightarrow$ & $\text{HTP}_{311}^\updownarrow$ & $\text{P}_{010}$  \\
    \hline 
	$y_n$ & 3.484 & 7.079 &  7.095 & 7.122 & 7.154 & 7.292 & 7.301 \\
\hline
\end{tabular}
\caption{Magnetoquasistatic eigenvalues $y_n$ of an ring with minor radius $r$ and major radius $R=3r$, We assumed $l_c = 3 R$.}
\label{tab:MQStorus}
\end{table}

{\bf Ring.} Eventually,  we investigate a dielectric ring (solid torus) with minor radius $r$ and major radius $R=3r$, whose boundary has a different genus with respect to the object that we have investigated so far. Similar rod-ring shaped dielectric resonators have been investigated in the context of antennas,  e.g. \cite{ong2004rod}. In Fig. \ref{fig:Eigenvalues} (d) we show the first 100 magnetostatic eigenvalues, while  in Tab. \ref{tab:MQStorus} we list the values only of the low-order ones.

We catalog these modes using a {\it magnetic} coordinate system \cite{wesson2011tokamaks} where $\bf \hat{\phi}$, $\bf \hat{r}$, $\bf \hat{\theta}$  are the toroidal, radial and poloidal directions, respectively.  We describe the number of oscillation of the displacement current density modes along the  $\bf \hat{\phi}$, $\bf \hat{r}$, $\bf \hat{\theta}$ direction by the toroidal $t$,  radial $r$, and poloidal $p$ numbers. The modes which are invariant along the toroidal direction ($t=0$) are denoted either as toroidal $\text{T}_{0rp}$  if the displacement current is directed along $\bf \hat{\phi}$ or poloidal $\text{P}_{0rp}$ if the displacement current is directed along $\bf \hat{\theta}$.  We denote all the remaining modes as {\it hybrid toroidal-poloidal} modes $\text{HTP}_{mrp}$.

In Fig. \ref{fig:ModesTorus} we show the displacement current modes of the first seven non-degenerate eigenvalues. The first mode, shown in Fig \ref{fig:ModesTorus} (a),  is the fundamental toroidal mode $\text{T}_{010}$: its  magnetostatic energy  is significantly lower with respect to the remaining modes. The fundamental poloidal mode  $\text{P}_{010}$ is  the seventh non-generate mode, shown in Fig. \ref{fig:ModesTorus} (g). The remaining modes in Fig. \ref{fig:ModesTorus} are hybrid. Specifically, the modes in Fig. \ref{fig:ModesTorus} (c)-(c') and  (d)-(d') are characterized by the same numbers $\left(t,r,p \right)=\left(1,1,1\right)$: we distinguish them as up-down $\text{HTP}_{111}^\updownarrow$ (c)-(c'), and inboard-outboard $\text{HTP}_{111}^\leftrightarrow$  (d)-(d'). Similarly, the modes in Fig. \ref{fig:ModesTorus} (b)-(b') and  (e)-(e') are denoted as   $\text{HTP}_{211}^\updownarrow$ and  $\text{HTP}_{211}^\leftrightarrow$. Next, we show in Fig. \ref{fig:ModesTorus} (f)  $\text{HTP}_{311}^\updownarrow$. 

\section{Conclusions}
There exist two mechanisms through which a small non-magnetic homogeneous object may resonate \cite{SupMat}. The first is the electroquasistatic resonance \cite{Mayergoyz:03} where the induced electric charge plays a central role. These resonances are connected to the eigenvalues of the electrostatic integral operator that gives the electrostatic field as a function of the charge density. They physically arise from the interplay between the  energy stored in the electric field and the kinetic energy of the electrons.  Each resonance is characterized by a negative eigenpermittivity, which is size-independent.  The eigenpermittivities constitute a countable infinite set, and accumulate at the point $-1$. The induced current density fields are both curl-free and div-free within the particle, but have non-vanishing (and discontinuous) normal components to the particle surface.

The second mechanism is the magnetoquasistatic resonance, described in this paper, where the  displacement current density field is the main player.  These resonances are connected to the eigenvalues of a  magnetostatic integral operator that gives the vector potential as a function of the current density. They arise from the interplay between the polarization energy stored in the dielectric and the energy stored in the magnetic field. These resonances are only possible for positive permittivity. For any given shape, the resonance frequencies are inversely proportional to the characteristic size of the object, and inversely proportional to the square root of the permittivity.  They are an infinite but countable set accumulating at $+\infty$.  The induced current density fields have a non-zero curl within the particle, but are div-free and have a vanishing normal component on the particle surface.

The applicability of the magnetoquasistatic approximation can be extended to accurately describe the radiation damping of the modes and the frequency-shift due to the finite particle particle size by using perturbation techniques.

\appendix
\section{Solution of the electromagnetic scattering in the quasi-static limit}
\label{sec:A}
Let us consider an isotropic and homogeneous material occupying a volume  $\Omega$, which is bounded by a closed surface $\partial \Omega$   with outward-pointing normal $\n$ . The medium is nonmagnetic with relative permittivity $\varepsilon_R$, surrounded by vacuum. 
The object is illuminated by a time harmonic electromagnetic field incoming from infinity $\text{Re} \left\{ \mathbf{E}_{inc} \left( \mathbf{r} \right) e^{i \omega t } \right\}$, where $\omega$ is the angular frequency. Here, we derive the behavior of the electromagnetic scattering in the quasi-static limit starting from the full wave model.

The scattering problem is formulated by considering as unknown the effective current density field  $\mathbf{J} = \mathbf{J} \left( \mathbf{r} \right)$  induced in the body (which particularizes into conduction current in metals at frequencies below interband transitions, polarization current in dielectrics, sum of conduction and polarization currents in metals in frequency ranges where interband transitions occur). 
We have $\mathbf{J} = i \omega \varepsilon_0 \chi \mathbf{E}$  where $\mathbf{E} = \mathbf{E} \left( \mathbf{r} \right)$  is the total electric field (induced + incident) and  $\chi = \left( \varepsilon_R  - 1 \right)$ is the electric susceptibility. Both the vector fields $\mathbf{E}$ and $\mathbf{J}$  are div-free in $\Omega$  due to the homogeneity and isotropy of the material.
The current density $\mathbf{J}$  is governed by the full wave volume integral equation \cite{van2007electromagnetic,Schaubert:84,jin2011theory,rubinacci2006broadband,miano2010numerical}:
\begin{multline}
   \frac{\Jr}{i \omega \varepsilon_0 \chi} = - \frac{1}{i \omega \varepsilon_0} \nabla \oiint_{\partial \Omega} \g \Jr \cdot \n \left( \mathbf{r}' \right) dS \\- i \omega \mu_0 \iiint_V \g \Jr dV + \Ei \quad \forall \mathbf{r} \in \Omega,
   \label{eq:VIE} 
\end{multline}
where $\varepsilon_0$  is the vacuum permittivity, $\mu_0$  is the vacuum permeability,  $g \left( \mathbf{r} \right) = e^{-i k_0 r} / 4 \pi r $  is the Green function in vacuum, $k_0 = \omega/ c_0$   and $c_0 = 1 / \sqrt{\varepsilon_0 \mu_0}$.
The surface integral represents the contribution of the scalar potential to the induced electric field and the volume integral represents the contribution of the vector potential. We introduce the dimensionless size parameter $ x = \omega l_c / c_0$  where $l_c$  is a characteristic linear length of the region  $\Omega$. 
Then, equation \ref{eq:VIE} is rewritten as
\begin{equation}
  \frac{ \mathbf{J} \left( \rt \right) }{\chi} - \mathcal{L} \left\{ \mathbf{J} \right\} \left( \rt \right) = i \omega \varepsilon_0 \mathbf{E}_{inc} \left( \rt \right) \quad \forall \rt\in \tilde{\Omega} 
 \label{eq:VIEnorm}
\end{equation}
where $\rt = \mathbf{r}/l_c$,
\begin{multline}
 \mathcal{L} \left\{ \mathbf{W} \right\} \left( \rt \right) =
  - \tnabla \oiint_{\partial \tilde{\Omega}} \gao \Wn d\tilde{S}' \\\ + x^2 \iiint_{\tilde{\Omega}} \gao \W d\tilde{V}'
\end{multline}
$\tilde{\Omega}$ is the scaled domain,  $\tilde \nabla$  is the scaled gradient operator and $W_n = {\bf W} \cdot \n$. In the quasi-static limit ($x \rightarrow 0$) the operator $\mathcal{L}$ has the following expression up to the third order in x:
\begin{widetext}
\begin{equation}
  \mathcal{L} \left\{ \mathbf{W} \right\} \left( \rt \right) =  - \tilde{\nabla} \oiint_{\partial \tilde{\Omega}} \go \left( 1 - i x \left| \rt - \rt' \right| + \frac{x^2}{2} \left| \rt - \rt' \right|^2 \right) \Wn d\tilde{S}' \\ + x^2 \iiint_{\tilde{\Omega}} \go \W d\tilde{V}' + O( x^3) 
\end{equation}
\end{widetext}
$\text{as} \, x \rightarrow 0$, where 
\begin{equation}
  g_0 \left( \mathbf{r} \right) = \frac{1}{4 \pi r}
\end{equation}
is the static Green function for the vacuum. 

We now study the solution of equation \ref{eq:VIEnorm} in the quasi-static limit $x \rightarrow 0 $. To achieve this purpose we introduce a complete basis for representing the unknown, which is obtained by joining two orthogonal sets. The first set $\left\{ \Whp \right\}$  is given by the solution of the eigenvalue problem
\begin{equation}
\mathcal{L}_e \left\{ \Whp \right\} = \frac{1}{\ghp} \Whp,
\end{equation}
where $\mathcal{L}_e$  is the electrostatic integral operator
\begin{equation}
  \mathcal{L}_e \left\{ \mathbf{W} \right\} = - \tilde{\nabla} \oiint_{ \partial \tilde{\Omega}} \go \Wn d\tilde{S}'.
\end{equation}
The eigenfunctions $\left\{ \Whp  \right\}$ are both div-free and curl-free  in the limit $x \rightarrow 0$ in $\Omega$, but they have non-vanishing normal components to  $\partial \Omega$. Since  $\mathcal{L}_e$  is Hermitian and definite negative the eigenvalues  $\ghp$ are real and negative, and the eigenfunctions are orthonormal according to the scalar product 
\begin{equation}
 \langle \mathbf{A}, \mathbf{B} \rangle = \iiint_\Omega \mathbf{A} \cdot \mathbf{B} \, dV.
 \label{eq:ScalarProduct}
\end{equation}
Both the eigenfunctions $\left\{ \mathbf{W}_h^\parallel \right\}$  and the eigenvalues $\left\{ \gamma_h^\parallel \right\}$  do not depend on the size of object, but only on its shape. 
However, the set $\left\{ \Whp \right\}$  is not sufficient to represent the vector space of square integrable div-free functions in $\Omega$. To complete the basis it is sufficient to add to  $\left\{ \Whp \right\}$ the set of solenoidal vector fields $\left\{ \Who \right\}$  with vanishing normal components to  $\partial \Omega$ and that are solution of the eigenvalue problem in weak form
\begin{equation}
\mathcal{L}_m \left\{ \Who \right\} = \frac{1}{\ghp} \Who
\label{eq:MQSproblem}
\end{equation}
where $\mathcal{L}_m$  is the {\it magnetostatic integral operator}
\begin{equation}
\mathcal{L}_m \left\{  \Who \right\} \left( \rt \right)  = \iiint_{\tilde{\Omega}} \go \Who \left( \rt' \right)  d\tilde{V}'.
\label{eq:MQSoperator}
\end{equation}
The eigenfunctions $\left\{ \mathbf{W}_h^\perp \right\}$  are div-free in $\Omega$  and have vanishing normal components to $\partial \Omega$, but their curl in $\Omega$  is different from zero. Since  $\mathcal{L}_m$ is Hermitian and definite positive the eigenvalues $\gho$ are real and positive, and the eigenfunctions are orthonormal according to the scalar product \ref{eq:ScalarProduct}. Both the eigenfunctions $\left\{ \Who \right\}$  and the eigenvalues   do not depend on the size of object, but only on its shape. Furthermore, the set of eigenfunctions $\left\{ \Whp \right\}$ is orthogonal to set of eigenfunctions  $\left\{ \Who \right\}$ . 
The union of the two sets $\left\{ \Whp \right\}$  and $\left\{ \Who \right\}$  is a complete basis for the vector space of square integrable div-free vector fields in  $\Omega$. Therefore, we represent the unknown current density $\mathbf{J}$  as
\begin{equation}
   \label{eq:expansion}
   \mathbf{J} \left( \rt \right) = \sum_h \left( I_h^\parallel \Whp \left( \rt \right)  + I_h^\perp \Who \left( \rt \right)  \right).
\end{equation}
We substitute expression \ref{eq:expansion} into equation \ref{eq:VIEnorm} and we obtain to the leading order in  $x$  the following expressions for the expansion coefficients  $I_h^\perp$ and  $I_h^\perp$:
\begin{equation}
\begin{aligned}
  I_h^\parallel &= \frac{\ghp}{\ghp - \chi } i \omega \chi \varepsilon_0 \langle \Whp, \Ei \rangle \\
    I_h^\perp &= \frac{\gho}{\gho - x^2 \chi } i \omega \chi \varepsilon_0 \langle \Who, \Ei \rangle
\end{aligned}
\label{eq:Coefficients}
\end{equation}
Expression \ref{eq:expansion} with \ref{eq:Coefficients} is the solution of the scattering problem \ref{eq:VIEnorm} in the quasi-static limit  $x \rightarrow 0$.
The eigenfunction  $\Whp$ may be resonantly exited when $\text{Re} \left\{ \chi \right\} = \gho$  and the eigenfunction $\Who$  may be resonantly exited when  $\text{Re}\left\{\chi\right\} = \ghp/x^2$. For these reasons the sets    $\left\{ \Whp \right\}$ and $\left\{ \Whp \right\}$  can be interpreted as the current modes of the body in the quasi-static limit  $x \rightarrow 0$, and  $\left\{ \ghp \right\}$ and $\left\{ \gho  \right\}$  can be interpreted as the corresponding eigen-susceptibilities. Thus, we call the eigenfunctions $\left\{ \Whp \right\}$  electroquasistatic current modes and the eigenfunctions $\left\{ \Whp \right\}$   magnetoquasistatic current modes. Since the eigen-susceptibilities $\left\{ \ghp  \right\}$  are all negative the electroquasistatic modes can be resonantly excited only in metals (surface plasmons). On the contrary, the eigen-susceptibilities  $\left\{ \ghp/x^2 \right\}$  are all positive, therefore the magnetoquasistatic modes can be resonantly excited only in dielectrics.
The scalar products $ \langle \Whp, \Ei \rangle$ and $\langle \Who, \Ei \rangle$ in Eq. \ref{eq:Coefficients} describe the coupling of the electroquasistatic and magnetoquasistatic current modes with the external excitation.

The magnetoquasistatic eigenvalue problem \ref{eq:MQSproblem} is the problem \ref{eq:Auxiliary} of the main manuscript where $\gamma^\perp$  is replaced by  $y^2$.  Since $\displaystyle\min_{h\in\mathbb{N}}{\ghp} \approx 1 $  and  $x \ll 1$  the magnetoquasistatic resonances occur in dielectrics with high permittivity, hence the resonance frequency of the $h^{th}$ magnetoquasistatic mode is  $\omega_h \approx \left( \frac{c_0}{ l_c \sqrt{\varepsilon_r}} \right) \sqrt{\gho}$: this is the resonant condition \ref{eq:ResonantFreq} of the main manuscript (where $\gamma^\perp$  is replaced by $y^2$ ). Moreover, the problem \ref{eq:MQSproblem} can be rewritten in differential form as in the following:
\begin{equation}
\tnabla \times \tnabla \times \mathbf{A}^\perp = \gamma^\perp \mathbf{A}^\perp \tilde{\Pi} \left( \rt' \right)
\label{eq:DiffProblem}
\end{equation}
where $\tnabla \times$ is the scaled curl, $\Ao \left( \mathbf{r} \right) = \mathcal{L}_m \left\{ \Wo \right\} \left( {\bf r} \right)$  for $\rt \in \tOmega$  and $\tPi \left( \rt \right)=1$ for $\rt \in \tilde{\Omega}$ and $0$ for $\rt \notin \tilde{\Omega}$. Note that at resonance we have  $\gho=l_c^2 \beta^2$. This validates the magnetoquasistatic model defined by Eqs. \ref{eq:RotA}, \ref{eq:RotH}, and \ref{eq:J} of the main manuscript.

\section{Energy balance in the magnetoquasistatic resonances}
\label{sec:B}
In the electroquasistatic resonances of metals below the interband transitions the energy oscillates back and forth between the kinetic energy of the conduction electrons and the Coulomb potential energy arising from the surface charges on the surface of the metal. In the magnetoquasistatic resonances in dielectrics the energy oscillates back and forth between the polarization energy of the dielectric and the magnetic energy. Indeed we now show that in these resonances the energy stored in the magnetic field is balanced by the energy stored in the dielectric in the form of polarization energy. By assuming that the dielectric susceptibility is real, the resonance frequency $\omega_h$ for the magnetoquasistatic current mode $\Who$  is given by the condition
\begin{equation}
   \frac{\omega_h^2 l_c^2}{c_0^2} \chi = \gamma_h^\perp.
   \label{eq:Ares}
\end{equation}
The eigenvalue $\gamma_h$ is related to the time average of the magnetic energy. Indeed, we have
\begin{equation}
   \gho = \frac{\left\| \tnabla \times \Aho \right\|_{\mathbb{R}^3}^{2}}{\left\|  \Aho \right\|_{\tilde{\Omega}}^{2}}
\label{eq:Aenergy}
\end{equation}
where $\mathbf{A}_h^\perp$  is the magnetic vector potential generated by the current mode  $\Who$ and $\tnabla \times$ is the scaled curl. By combining Eqs. \ref{eq:Ares} and \ref{eq:Aenergy} we obtain
\begin{equation}
  \frac{1}{2\mu_0} \left\| \tnabla \times \mathbf{A}_h^\perp \right\|_{\mathbb{R}^3}^2 = \frac{\varepsilon_0 \chi}{2} \left\| \omega_h {\bf A}_h^\perp \right\|^2_\Omega.
\end{equation}
The term on the left hand side is the energy stored in the magnetic field associated to the current mode  $\Who$ while term on the right hand side is the energy stored in the dielectric, in the form of polarization energy, at the resonance frequency $\omega_h$.

\section{Numerical Model}
\label{sec:C}
Equation \ref{eq:Auxiliary} can be discretized by drawing on the standard repertoire of computational electromagnetics for Volume Integral Equations \cite{Schaubert:84,jin2011theory,van2007electromagnetic}.  The unknown of the magnetoquasistatic problem, i.e. the displacement current density field $\Jp$, belongs to the functional space $\mathcal{J}_{L}$ \cite{bossavit1998computational}:
\begin{equation*}
 \mathcal{J}_{L}  =\left\{  \mathbf{w}  \in H \left(  \operatorname{div} 
,\Omega\right)  |\nabla\cdot\mathbf{w}=0\text{ in }\Omega , %
\mathbf{w}\cdot\mathbf{\hat{n}}=0 \text{ on } \partial\Omega\right\}
\end{equation*}
We obtain the discretization of Eq. \ref{eq:Auxiliary} by representing $\Jp$ in terms of $N_L$ loop shape functions ${\bf w}^L_k$ 
\begin{equation}
\Jp = \sum_{k=1}^{N_{L}}I_{k}^{L}\mathbf{w}_{k}^\text{L} 
\label{eq:representation}
\end{equation}
Each function $\mathbf{w}_{k}^\text{L}$ is associated to the $k$-th edge of the finite element discretization of the volume $\Omega$. It is defined as the curl of the $k$-th edge-element shape functions \cite{bossavit1998computational}  $\mathbf{N}_{k}$:
\[
\mathbf{w}_{k}^\text{L}\left(  \mathbf{r}\right)  =\nabla\times\mathbf{N}%
_{k}\left(  \mathbf{r}\right).
\]
The discrete generalized eigenvalue problem is obtained by substituting the representation (\ref{eq:representation}) into Eq. (\ref{eq:Auxiliary}) and applying the Galerkin method, projecting along the loop shape functions:
\begin{equation}
y^2 \mathbf{ L} \, \mathbf{I} = \mathbf{R} \, \mathbf{I}.
\label{eq:NumEigPro}%
\end{equation}
The matrices $\mathbf{R}$ and $\mathbf{L}$ are associated to the l.h.s. and r.h.s. of Eq. \ref{eq:Auxiliary}, respectively. The generic occurrences of these matrices are:
\begin{align*}
  \mathbf{R}  _{pq}  & =\int_{\tilde{\Omega}}\mathbf{w}%
_{p}^\text{L}\left(  \mathbf{\tilde{r}}\right) \cdot \mathbf{w}_{q}^\text{L} \left(  \mathbf{\tilde{r}}\right) d \tilde{V}\\
 \mathbf{L}  _{pq}  & = \int_{\tilde{\Omega}}%
\int_{\tilde{\Omega}}\mathbf{w}_{p}^\text{L}\left(  \mathbf{\tilde{r}}\right)  \cdot
\mathbf{w}_{q}^\text{L}\left( \mathbf{\tilde{r}}^{\prime}\right)  g_0\left(
\mathbf{\tilde{r}-\tilde{r}}^{\prime}\right)  \text{d}\tilde{V}\text{d}\tilde{V}^{\prime}.
\end{align*}
Eventually, the problem \ref{eq:NumEigPro} is reduced to a standard symmetric eigenvalue problem by exploiting the LAPACK \cite{laug} routine DSYGST, then all eigenvalues and eigenvectors of the resulting real symmetric matrix are computed through the routine DSYEV. In Tab. \ref{tab:Mesh} we provide details about the number of nodes, elements, and edges of the hexahedral meshes used in the calculation.

\begin{table}[h!]
\centering
\begin{tabular}{|c||ccc|}
\hline
 object & $N_{node}$ & $N_{elem}$ & $N_{edge}$ \\
\hline
sphere & 6527    &  6048  &  11665  \\
finite-size cylinder & 6060  & 5148 & 9433  \\
triangular prism & 2520   &  2025  & 3592  \\
ring & 29056  & 26112 &  49409  \\
\hline
\end{tabular}
\caption{Details on the meshes used in the numerical calculation of the eigenvalues.}
\label{tab:Mesh}
\end{table}

\end{document}